\pdfoutput=1

\documentclass[11pt]{article}

\usepackage[]{ACL2023}
\usepackage{float}
\usepackage{times}
\usepackage{rotating}
\usepackage{soul}
\usepackage{url}
\usepackage{graphicx}
\usepackage{amsmath}
\usepackage{amsthm}
\usepackage{booktabs}
\usepackage{algorithm}
\usepackage{algorithmic}
\usepackage[switch]{lineno}
\usepackage{pifont}
\usepackage{multirow}
\usepackage{natbib}
\usepackage{tcolorbox}
\usepackage{enumitem}
\usepackage{setspace}
\usepackage{cleveref}
\usepackage{xcolor}
\usepackage[font=small,labelfont=bf]{caption}
\usepackage{array,ragged2e}
\usepackage{subcaption}
\usepackage{listings}
\usepackage{makecell}
\usepackage{forest}
\usepackage{hyperref}
\usepackage{tikz}
\usepackage{longtable}
\usepackage{tablefootnote}
\usepackage{graphics}
\newcommand{\cmark}{\ding{51}}%
\definecolor{mygreen}{RGB}{0,150,0}
\definecolor{myred}{RGB}{255,0,0}
\definecolor{paired-light-blue}{RGB}{198, 219, 239}
\definecolor{paired-dark-blue}{RGB}{49, 130, 188}
\definecolor{paired-light-orange}{RGB}{251, 208, 162}
\definecolor{paired-dark-orange}{RGB}{230, 85, 12}
\definecolor{paired-light-green}{RGB}{199, 233, 193}
\definecolor{paired-dark-green}{RGB}{49, 163, 83}
\definecolor{paired-light-purple}{RGB}{218, 218, 235}
\definecolor{paired-dark-purple}{RGB}{117, 107, 176}
\definecolor{paired-light-gray}{RGB}{217, 217, 217}
\definecolor{paired-dark-gray}{RGB}{99, 99, 99}
\definecolor{paired-light-pink}{RGB}{222, 158, 214}
\definecolor{paired-dark-pink}{RGB}{123, 65, 115}
\definecolor{paired-light-red}{RGB}{231, 150, 156}
\definecolor{paired-dark-red}{RGB}{131, 60, 56}
\definecolor{paired-light-yellow}{RGB}{231, 204, 149}
\definecolor{paired-dark-yellow}{RGB}{141, 109, 49}
\definecolor{bg1}{HTML}{FF9966}
\definecolor{bg2}{HTML}{CCE5FF}
\definecolor{bg3}{HTML}{FFCC99}
\definecolor{bg4}{HTML}{FFC107}
\definecolor{bg5}{HTML}{FFCCCC}
\definecolor{bg6}{HTML}{D5E8D4}
\definecolor{bg7}{HTML}{eeeeee}
\definecolor{bg8}{HTML}{cdeb8b}
\definecolor{bg9}{HTML}{dae8fc}
\definecolor{bg10}{HTML}{a2e6eb}
\definecolor{bg31}{HTML}{FFCDD2}
\definecolor{bg32}{HTML}{F8BBD0}
\definecolor{bg33}{HTML}{E1BEE7}
\definecolor{bg34}{HTML}{D7CCC8}
\definecolor{bg35}{HTML}{B2DFDB}
\definecolor{bg36}{HTML}{A5D6A7}
\definecolor{bg37}{HTML}{FFF9C4}
\definecolor{bg38}{HTML}{FFECB3}
\definecolor{bg111}{HTML}{CB6843}
\definecolor{bg112}{HTML}{D77C5C}
\definecolor{bg113}{HTML}{E28E6E}
\definecolor{bg114}{HTML}{E89F7D}
\definecolor{bg115}{HTML}{EDAE8A}
\definecolor{bg116}{HTML}{F0BA95}
\definecolor{bg117}{HTML}{F3C29F}
\definecolor{bg118}{HTML}{F6CCAA}
\definecolor{bg119}{HTML}{F8D5B3}
\definecolor{bg120}{HTML}{FADCBD}
\definecolor{bg121}{HTML}{FCE6C7}
\definecolor{bg39}{HTML}{FFE0B2}
\definecolor{bg40}{HTML}{3CB371}
\definecolor{bg43}{HTML}{ffe5d9}
\definecolor{bg15}{HTML}{7FFFD4}
\definecolor{bg17}{HTML}{F0FFFF}
\definecolor{bg18}{HTML}{F5FFFA}
\definecolor{bg19}{HTML}{F8F8FF}
\definecolor{bg20}{HTML}{FFFFFF}
\definecolor{bg21}{HTML}{E1F5FE}
\definecolor{bg22}{HTML}{B3E5FC}
\definecolor{bg23}{HTML}{81D4FA}
\definecolor{bg24}{HTML}{4FC3F7}
\definecolor{bg25}{HTML}{29B6F6}
\definecolor{bg26}{HTML}{03A9F4}
\definecolor{bg27}{HTML}{039BE5}
\definecolor{bg28}{HTML}{0288D1}
\definecolor{bg29}{HTML}{0277BD}
\definecolor{bg30}{HTML}{01579B}
\definecolor{bg16}{HTML}{FFCC99}
\definecolor{pg51}{HTML}{E8F5E9}
\definecolor{pg52}{HTML}{C8E6C9}
\definecolor{pg53}{HTML}{B9F6CA}
\definecolor{pg54}{HTML}{A9DFBF}
\definecolor{pg55}{HTML}{BCF5A6}
\definecolor{pg56}{HTML}{BEF1CE}
\definecolor{pg57}{HTML}{CEF6EC}
\definecolor{pg58}{HTML}{B7F0B1}
\definecolor{pg59}{HTML}{B1F2B5}
\definecolor{pg60}{HTML}{9DF3C4}
\definecolor{pg61}{HTML}{DEF7E0}
\definecolor{pg62}{HTML}{E8F8DC}
\definecolor{pg63}{HTML}{EBF7E7}
\definecolor{pg64}{HTML}{F0FDF4}
\definecolor{pg65}{HTML}{F1FEE7}
\definecolor{pg66}{HTML}{F7FFF6}
\definecolor{pg67}{HTML}{FCFFE7}
\definecolor{pg68}{HTML}{F4FFD2}
\definecolor{pg69}{HTML}{EEFFE2}
\definecolor{pg70}{HTML}{E3FDF5}
\definecolor{connect-color}{RGB}{0,0,0}
\definecolor{middle-color}{RGB}{255,255,255}
\definecolor{leaf-color}{RGB}{173,216,230}
\definecolor{line-color}{RGB}{25,25,112}

\usepackage{multirow}

\tikzset{
    root style/.style={
        draw,
        rounded corners,
        fill=blue!30, 
        align=center,
        font=\bfseries
    },
    child style/.style={
        draw,
        rounded corners,
        fill=green!30, 
        align=center,
        font=\bfseries
    },
    grandchild style/.style={
        draw,
        rounded corners,
        fill=red!30, 
        align=center,
        font=\bfseries
    }
}
\tikzset{
  my-box/.style={
    rectangle,
    draw=hidden-draw,
    rounded corners,
    text opacity=1,
    minimum height=1.5em,
    minimum width=40em,
    inner sep=2pt,
    align=center,
    line width=0.8pt,
  },
  leaf/.style={
    my-box,
    minimum height=1.5em,
    text=black,
    align=center,
    font=\normalsize,
    inner xsep=2pt,
    inner ysep=4pt,
    line width=0.8pt,
  }
}

\title{Breaking Down the Defenses: A Comparative Survey of Attacks on Large Language Models}

\author{
    \textbf{Arijit Ghosh Chowdhury}$^1$, \textbf{Md Mofijul Islam}$^{3*}$, \textbf{Vaibhav Kumar}$^5$, \\ \textbf{Faysal Hossain Shezan}$^4$, \textbf{Vaibhav Kumar}$^6$, \textbf{Vinija Jain}$^2$, \textbf{Aman Chadha}$^{2,3}$\thanks{\,\,\,Work does not relate to position at Amazon.}
    \\
    $^1$University of Illinois Urbana-Champaign\\$^2$Stanford University\quad$^3$Amazon GenAI $^4$University of Texas at Arlington\\
    $^5$University of California, Los Angeles 
    $^6$Georgia Institute of Technology\\
    \texttt{arijit10@gmail.com, mi8uu@virginia.edu, vaibhavk@ucla.edu} \\
    \texttt{faysal.shezan@uta.edu, vaibhavk@gatech.edu, hi@vinija.ai, hi@aman.ai}
}

\begin{document}
\maketitle

\begin{abstract}
    
Large Language Models (LLMs) have become a cornerstone in the field of Natural Language Processing (NLP), offering transformative capabilities in understanding and generating human-like text. However, with their rising prominence, the security and vulnerability aspects of these models have garnered significant attention. This paper presents a comprehensive survey of the various forms of attacks targeting LLMs, discussing the nature and mechanisms of these attacks, their potential impacts, and current defense strategies. We delve into topics such as adversarial attacks that aim to manipulate model outputs, data poisoning that affects model training, and privacy concerns related to training data exploitation. The paper also explores the effectiveness of different attack methodologies, the resilience of LLMs against these attacks, and the implications for model integrity and user trust. By examining the latest research, we provide insights into the current landscape of LLM vulnerabilities and defense mechanisms. Our objective is to offer a nuanced understanding of LLM attacks, foster awareness within the AI community, and inspire robust solutions to mitigate these risks in future developments.
\end{abstract}
\section{Introduction}

The emergence of artificial intelligence has marked a significant transformation in Natural Language Processing through the introduction of large language models (LLMs) enabling unprecedented advances in language comprehension, generation, and translation ~\cite{zhao2023survey, naveed2023comprehensive, achiam2023gpt}. Despite their transformative impact, LLMs have become susceptible to a variety of sophisticated attacks, posing significant challenges to their integrity and reliability ~\cite{yao2023survey, liu2023summary}. This survey paper provides a comprehensive examination of the attacks targeting LLMs, elucidating their mechanisms, consequences, and the fast evolving threat landscape.

The significance of investigating attacks on LLMs lies in their extensive integration across various sectors and their consequential societal ramifications ~\cite{eloundou2023gpts}. LLMs are instrumental in applications ranging from automated customer support to sophisticated content creation. Therefore, understanding their vulnerabilities is imperative for ensuring the security and trustworthiness of AI-driven systems ~\cite{amodei2016concrete, hendrycks2023overview}. This paper categorizes the spectrum of attacks, based on access to model weights and attack vectors, each presenting distinct challenges and requiring specific attention.

Additionally, the methodologies employed in executing these attacks are dissected, offering insights into the adversarial techniques utilized to exploit LLM vulnerabilities. While acknowledging the limitations of current defense mechanisms, the paper also proposes potential avenues for future research in enhancing LLM security.

We summarize the major contributions of our work as follows:
\begin{tcolorbox}[colback=blue!5!white,colframe=blue!75!black,title={\textbf{\textsc{{Our Contributions}}}}]
\begin{itemize}
 [leftmargin=1mm]
 \setlength\itemsep{0em}
 \begin{spacing}{0.85}
 \vspace{-1mm}
\item[\ding{224}] {\footnotesize 
     {\fontfamily{phv}\fontsize{8}{9}\selectfont We propose a novel taxonomy of attacks on LLMs,
which can help researchers to better understand the research landscape and fnd their areas of interest.}}
\vspace{-1mm}
\item[\ding{224}] {\footnotesize 
     {\fontfamily{phv}\fontsize{8}{9}\selectfont We present existing attack and mitigation approaches in detail, discussing key implementation details.}}
\vspace{-1mm}
\item[\ding{224}] {\footnotesize 
     {\fontfamily{phv}\fontsize{8}{9}\selectfont We discuss important challenges, highlighting promising directions for future research.}}
\vspace{-5.5mm}    
\end{spacing}    
 \end{itemize}
\end{tcolorbox}







\section{Exploring LLM Security: White and Black Box Attacks}

This section delves into the security challenges of LLMs from both white box and black box perspectives. It highlights the importance of understanding and protecting LLMs against complex security threats.

\subsection{White Box}

These attacks exploit full access to the LLM's architecture, training data, and algorithms, enabling attackers to extract sensitive information, manipulate outputs, or insert malicious code. \citet{shayegani2023survey} discusses whitebox attacks, highlighting how this access permits crafting adversarial inputs to alter outputs or impair performance. The study covers various attack strategies, such as context contamination and prompt injection, aimed at manipulating LLMs for specific outputs or reducing their quality.

Separately, \citet{li2023privacy} examines privacy concerns in LLMs, emphasizing the importance of protecting personal information in the face of evolving AI technologies. They discuss the privacy risks associated with training and inference data, highlighting the critical need to analyze whitebox attacks for effective threat mitigation.

\subsection{Black Box}

These attacks exploit LLM vulnerabilities with limited knowledge of the model's internals, focusing on manipulating or degrading performance through the input-output interface. This approach, realistic in practical scenarios, poses risks such as sensitive data extraction, biased outputs, and diminished trust in AI. \citet{chao2023jailbreaking} illustrates black-box methods to ``jailbreak" LLMs like GPT-3.5 and GPT-4, with \citet{qi2023visual, yong2023low} exploring attacks on API-based models such as GPT-4 across various surfaces.


\section{LLM Attacks Taxonomy}
\definecolor{mygreen}{RGB}{0,150,0}
\definecolor{myred}{RGB}{255,0,0}

\newcolumntype{P}[1]{>{\RaggedRight\arraybackslash}p{#1}}


\definecolor{paired-light-blue}{RGB}{198, 219, 239}
\definecolor{paired-dark-blue}{RGB}{49, 130, 188}
\definecolor{paired-light-orange}{RGB}{251, 208, 162}
\definecolor{paired-dark-orange}{RGB}{230, 85, 12}
\definecolor{paired-light-green}{RGB}{199, 233, 193}
\definecolor{paired-dark-green}{RGB}{49, 163, 83}
\definecolor{paired-light-purple}{RGB}{218, 218, 235}
\definecolor{paired-dark-purple}{RGB}{117, 107, 176}
\definecolor{paired-light-gray}{RGB}{217, 217, 217}
\definecolor{paired-dark-gray}{RGB}{99, 99, 99}
\definecolor{paired-light-pink}{RGB}{222, 158, 214}
\definecolor{paired-dark-pink}{RGB}{123, 65, 115}
\definecolor{paired-light-red}{RGB}{231, 150, 156}
\definecolor{paired-dark-red}{RGB}{131, 60, 56}
\definecolor{paired-light-yellow}{RGB}{231, 204, 149}
\definecolor{paired-dark-yellow}{RGB}{141, 109, 49}

\definecolor{bg1}{HTML}{FF9966}
\definecolor{bg2}{HTML}{CCE5FF}
\definecolor{bg3}{HTML}{FFCC99}
\definecolor{bg4}{HTML}{FFC107}
\definecolor{bg5}{HTML}{FFCCCC}
\definecolor{bg6}{HTML}{D5E8D4}
\definecolor{bg7}{HTML}{eeeeee}
\definecolor{bg8}{HTML}{cdeb8b}
\definecolor{bg9}{HTML}{dae8fc}
\definecolor{bg10}{HTML}{a2e6eb}

\definecolor{bg31}{HTML}{FFCDD2} 

\definecolor{bg32}{HTML}{F8BBD0}

\definecolor{bg33}{HTML}{E1BEE7} 

\definecolor{bg34}{HTML}{D7CCC8} 

\definecolor{bg35}{HTML}{B2DFDB} 

\definecolor{bg36}{HTML}{A5D6A7} 

\definecolor{bg37}{HTML}{FFF9C4} 

\definecolor{bg38}{HTML}{FFECB3} 

\definecolor{bg111}{HTML}{CB6843}

\definecolor{bg112}{HTML}{D77C5C}

\definecolor{bg113}{HTML}{E28E6E}
\definecolor{bg114}{HTML}{E89F7D}
\definecolor{bg115}{HTML}{EDAE8A}
\definecolor{bg116}{HTML}{F0BA95}
\definecolor{bg117}{HTML}{F3C29F}
\definecolor{bg118}{HTML}{F6CCAA}
\definecolor{bg119}{HTML}{F8D5B3}
\definecolor{bg120}{HTML}{FADCBD}
\definecolor{bg121}{HTML}{FCE6C7}

\definecolor{bg39}{HTML}{FFE0B2} 

\definecolor{bg40}{HTML}{3CB371} 

\definecolor{bg43}{HTML}{ffe5d9}

\definecolor{bg15}{HTML}{7FFFD4}

\definecolor{bg17}{HTML}{F0FFFF}

\definecolor{bg18}{HTML}{F5FFFA}

\definecolor{bg19}{HTML}{F8F8FF}

\definecolor{bg20}{HTML}{FFFFFF}

\definecolor{bg21}{HTML}{E1F5FE}

\definecolor{bg22}{HTML}{B3E5FC}

\definecolor{bg23}{HTML}{81D4FA}

\definecolor{bg24}{HTML}{4FC3F7}

\definecolor{bg25}{HTML}{29B6F6}

\definecolor{bg26}{HTML}{03A9F4}

\definecolor{bg27}{HTML}{039BE5}

\definecolor{bg28}{HTML}{0288D1}

\definecolor{bg29}{HTML}{0277BD}

\definecolor{bg30}{HTML}{01579B}

\definecolor{bg16}{HTML}{FFCC99}

\definecolor{pg51}{HTML}{E8F5E9} 
\definecolor{pg52}{HTML}{C8E6C9} 
\definecolor{pg53}{HTML}{B9F6CA} 
\definecolor{pg54}{HTML}{A9DFBF} 
\definecolor{pg55}{HTML}{BCF5A6} 

\definecolor{pg56}{HTML}{BEF1CE} 
\definecolor{pg57}{HTML}{CEF6EC} 
\definecolor{pg58}{HTML}{B7F0B1} 
\definecolor{pg59}{HTML}{B1F2B5} 
\definecolor{pg60}{HTML}{9DF3C4} 

\definecolor{pg61}{HTML}{DEF7E0} 
\definecolor{pg62}{HTML}{E8F8DC} 

\definecolor{pg63}{HTML}{EBF7E7} 
\definecolor{pg64}{HTML}{F0FDF4} 

\definecolor{pg65}{HTML}{F1FEE7} 
\definecolor{pg66}{HTML}{F7FFF6} 
\definecolor{pg67}{HTML}{FCFFE7} 
\definecolor{pg68}{HTML}{F4FFD2} 
\definecolor{pg69}{HTML}{EEFFE2} 
\definecolor{pg70}{HTML}{E3FDF5} 

\definecolor{connect-color}{RGB}{0,0,0}
\definecolor{middle-color}{RGB}{255,255,255}
\definecolor{leaf-color}{RGB}{173,216,230}
\definecolor{line-color}{RGB}{25,25,112}



\definecolor{hidden-draw}{RGB}{20,68,106}
\definecolor{hidden-pink}{RGB}{255,245,247}
\definecolor{red}{RGB}{255,0,0}


\definecolor{hidden-draw}{RGB}{0,0,0}
\definecolor{hidden-pink}{RGB}{255,182,193}






\newcommand\vr[1]{\todo[author=VR,color=orange!40]{#1}}
\newcommand\vril[1]{\todo[author=VR,color=orange!40,inline]{#1}}

%
%

\tikzset{
    root style/.style={
        draw,
        rounded corners,
        fill=blue!30, 
        align=center,
        font=\bfseries
    },
    child style/.style={
        draw,
        rounded corners,
        fill=green!30, 
        align=center,
        font=\bfseries
    },
    grandchild style/.style={
        draw,
        rounded corners,
        fill=red!30, 
        align=center,
        font=\bfseries
    }
}

\tikzset{
  my-box/.style={
    rectangle,
    draw=hidden-draw,
    rounded corners,
    text opacity=1,
    minimum height=1.5em,
    minimum width=40em,
    inner sep=2pt,
    align=center,
    line width=0.8pt,
  },
  leaf/.style={
    my-box,
    minimum height=1.5em,
    text=black,
    align=center,
    font=\normalsize,
    inner xsep=2pt,
    inner ysep=4pt,
    line width=0.8pt,
  }
}

\begin{figure*}[ht!]
	\centering
	\resizebox{\textwidth}{!}{
		\begin{forest}
			for tree={
				grow=east,
				reversed=true,
				anchor=base west,
				parent anchor=east,
				child anchor=west,
				base=center,
				font=\large,
				rectangle,
				draw=hidden-draw,
				rounded corners,
				align=center,
				text centered,
				minimum width=5em,
				edge+={darkgray, line width=1pt},
				s sep=3pt,
				inner xsep=2pt,
				inner ysep=3pt,
				line width=0.8pt,
				ver/.style={
					rotate=90,
					child anchor=north,
					parent anchor=south,
					anchor=center
				},
			},
			where level=1{
				text width=15em,
				font=\normalsize,
			}{},
			where level=2{
				text width=14em,
				font=\normalsize,
			}{},
			where level=3{
				minimum width=10em,
				font=\normalsize,
			}{},
			where level=4{
				text width=26em,
				font=\normalsize,
			}{},
			where level=5{
				text width=20em,
				font=\normalsize,
			}{},
			[
				\textbf{Attacks on}\\
				\textbf{LLMs},
				for tree={fill=paired-light-red!70}
				[
					\textbf{Techniques to Attack},
					for tree={fill=paired-light-yellow!45}
					[
						\textbf{Jailbreaks}
						\hyperlink{jailbreaks}{\S3.1},
						for tree={fill=pg58},
						text width=13em
						[
							\textbf{\hyperlink{query}{Query-Based Jailbreaking}},
							for tree={fill=bg31},
							text width=13.2em
							[
								\textbf{PAIR} \cite{chao2023jailbreaking} \\
								\textbf{Competing Objectives} \cite{wei2023jailbroken},
								for tree={fill=bg2},
								leaf
							]
						]
						[
							\textbf{\hyperlink{prompt}{Prompt Engineering}},
							for tree={fill=bg31},
							text width=13.2em
							[
								\textbf{DeepInception} \cite{li2023deepinception}\\
								\textbf{ReNeLLM} \cite{ding2023wolf},
								for tree={fill=bg113},
								leaf
							]
						]
    					[
    						\textbf{\hyperlink{modal}{Cross-Modal Attacks}},
    						for tree={fill=bg31},
    						text width=13.2em
    						[
    							\textbf{Visual Adversarial Examples} \cite{qi2023visual} \\
    							\textbf{Low Resource JailBreaking} \cite{yong2023low},
    							for tree={fill=pg51},
    							leaf
    						]
    					]
    					[
    						\textbf{\hyperlink{automated}{Universal Attacks}},
    						for tree={fill=bg31},
    						text width=13.2em
    						[
    							\textbf{Universal Jailbreaks on Aligned LLMs} \cite{qi2023visual} \\
    							\textbf{Tree of Attacks} \cite{mehrotra2023tree} \\
    							\textbf{Persona Modulation} \cite{shah2023scalable},
    							for tree={fill=pg53},
    							leaf
    						]
    					]      
					]
				[
					\textbf{Prompt Injection} \hyperlink{prompt-injection}{\S3.2},
					for tree={fill=pg58},
					text width=13em
					[
						\textbf{\hyperlink{ObjectiveManipulation}{Objective Manipulation}},
						for tree={fill=bg33},
						text width=13.2em
						[
							\textbf{PromptInject} \cite{perez2022ignore} \\
							\textbf{Indirect Prompt Injection} \cite{abdelnabi2023not} \\
							\textbf{Propane} \cite{melamed2023propane},
							for tree={fill=orange},
							leaf
						]
					]
					[
						\textbf{\hyperlink{leaking}{Prompt Leaking}},
						for tree={fill=bg33},
						text width=13.2em
						[
							\textbf{HOUYI} \cite{liu2023prompt},
							for tree={fill=yellow},
							leaf
						]
					]
					[
						\textbf{\hyperlink{MCG}{Malicious Content Generation}},
						for tree={fill=bg33},
						text width=13.2em
						[
							\textbf{AutoDAN} \cite{liu2023prompt} \\
							\textbf{Prompt Packer} \cite{jiang2023prompt},
							for tree={fill=bg35},
							leaf
						]
					]
					[
						\textbf{\hyperlink{TrainingData}{Training Data Manipulation}},
						for tree={fill=bg33},
						text width=13.2em
						[
							\textbf{ProAttack} \cite{liu2023prompt},
							for tree={fill=bg1},
							leaf
						]
					]
				]
				[
					\textbf{Data Poisoning} \hyperlink{data-poisoning}{\S3.3},
					for tree={fill=pg58},
					text width=13em
					[
						\textbf{\hyperlink{PII}{PII Extraction}},
						for tree={fill=bg37},
						text width=13.2em
						[
							\textbf{Janus} \cite{chen2023janus} \\
							\textbf{PII Scrubing} \cite{lukas2023analyzing},
							for tree={fill=pg69},
							leaf
						]
					]
					[
						\textbf{\hyperlink{safety}{Safety Alignment Bypass}},
						for tree={fill=bg37},
						text width=13.2em
						[
							\textbf{Safety Degradation} \cite{qi2023fine} \\
							\textbf{Removing RLHF Protections} \cite{zhan2023removing} \\
							\textbf{Safety-Tuned Llamas} \cite{bianchi2023safety} \\
							\textbf{Forgetting Unsafe Examples} \cite{zhao2023learning},
							for tree={fill=paired-light-gray},
							leaf
						]
					]
					[
						\textbf{\hyperlink{backdoor}{Backdoor Attacks}},
						for tree={fill=bg37},
						text width=13.2em
						[
							\textbf{LOFT} \cite{shah2023loft} \\
							\textbf{AutoPoison} \cite{shu2023exploitability} \\
							\textbf{Backdoor Activation Attacks} \cite{bianchi2023safety} \\
							\textbf{Composite Backdoor Attacks} \cite{zhao2023learning},
							for tree={fill=paired-light-orange},
							leaf
						]
					]
				]    
				]
				[
					\textbf{Mitigation from Attacks} \hypertarget{mitigation}{},
					for tree={fill=bg22}
					[
						\textbf{Mitigation} \\ \textbf{Strategy} \S\ref{subsec:Mitigation Strategies},
						for tree={fill=bg16},
						text width=13em
						[
							\textbf{\hyperlink{IO}{Input/Output Censorship}},
							for tree={fill=pg70},
							text width=13.2em
							[
								\textbf{Baseline Defenses} \cite{jain2023baseline} \\
								\textbf{SmoothLLM} \cite{robey2023smoothllm} \\
								\textbf{Adversarial Prompt Shield} \cite{kim2023robust} \\
								\textbf{Llama Guard} \cite{inan2023llama} \\
								\textbf{NeMo-Guardrails} \cite{rebedea2023nemo} \\
								\textbf{Self-Examination} \cite{helbling2023llm},
								for tree={fill=cyan},
								leaf
							]
						]
						[
							\textbf{\hyperlink{MTFT}{Model Training/Fine-tuning}},
							for tree={fill=pg70},
							text width=13.2em
							[
								\textbf{Llama-2} \cite{touvron2023llama} \\
								\textbf{Context-Distillation} \cite{askell2021general},
								for tree={fill=paired-light-red},
								leaf
							]
						]
					]
				]
			]
		\end{forest}
	}
	\caption{Taxonomy of attacks and defenses on LLMs. We focus on prevalent methods across subthemes spanning jailbreaks, prompt injections and data poisoning. For Mitigation strategies, we divide papers into training based and moderation based efforts. All branches highlight key works that represent the themes.}
	\label{fig:lit_surv}
\end{figure*}
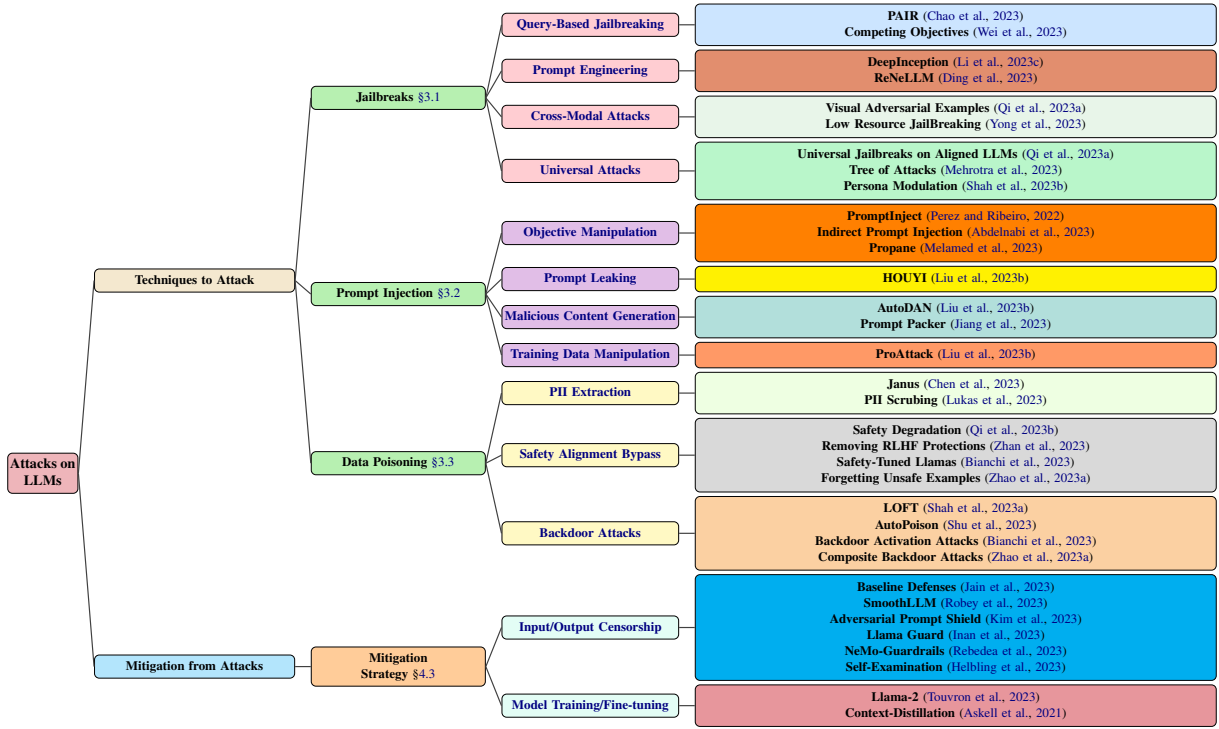

\subsection{Jailbreaks}
\hypertarget{jailbreaks}{}
This section delves into jailbreak attacks on LLMs, detailing strategies to exploit model vulnerabilities for unauthorized actions, underscoring the critical need for robust defense mechanisms.

\textbf{Refined Query-Based Jailbreaking:} \citet{chao2023jailbreaking} represent a strategic approach in jailbreaking, utilizing a minimal number of queries. This method doesn't just exploit simple model vulnerabilities but involves a nuanced understanding of the model's response mechanism, iteratively refining queries to probe and eventually bypass the model's defenses.
The success of this approach underscores a key vulnerability in LLMs: their predictability and manipulability through iterative, intelligent querying. This work introduces Prompt Automatic Iterative Refinement (PAIR), an algorithm designed to automate the generation of semantic jailbreaks for LLMs. PAIR works by using an attacker LLM to iteratively query a target LLM, refining a candidate jailbreak. This approach, more efficient than previous methods, requires fewer queries and can often produce a jailbreak in under twenty queries. PAIR demonstrates success in jailbreaking various LLMs, including GPT-3.5/4 and Vicuna, and is notable for its efficiency and interpretability, making the jailbreaks transferable to other LLMs.

\textbf{Sophisticated Prompt Engineering Techniques:} \citet{perez2022ignore} delve into the intricacies of LLMs' prompt processing capabilities. They demonstrate that embedding certain trigger words or phrases within prompts can effectively hijack the model's decision-making process, leading to the overriding of programmed ethical constraints.
\cite{ding2023wolf} focus on subtle, hard-to-detect jailbreaking methods using nested prompts. These findings reveal a critical shortcoming in the LLMs' content evaluation algorithms, suggesting the need for more complex, context-aware natural language processing that can discern and neutralize manipulative prompt structures. 

\textbf{Cross-Modal and Linguistic Attack Surfaces: } \citet{qi2023visual} reveals that LLMs are susceptible to multimodal inputs that combine text with visual cues. This approach takes advantage of the models' less robust processing of non-textual information. 
Similarly, \citet{yong2023low} exposes the heightened vulnerability of LLMs in processing low-resource languages. This indicates a significant gap in the models' linguistic coverage and comprehension, especially for languages with limited representation in training data. This work demonstrated that by translating unsafe English inputs into low-resource languages, it's possible to circumvent GPT-4's safety safeguards. 

\textbf{Universal and Automated Attack Strategies:} The development of universal and automated attack frameworks, as discussed in \cite{mehrotra2023tree} marks a pivotal advancement in jailbreaking techniques. These attacks involve appending specially chosen sequences of characters to a user's query, which can cause the system to provide unfiltered, potentially harmful responses. \citet{shah2023scalable} examine attacks leveraging the persona or style emulation capabilities of LLMs, introducing a new dimension to the attack strategies.

\subsection{Prompt Injection}
\label{sec:prompt-injection}
\hypertarget{prompt-injection}{}
This section outlines attacker strategies to manipulate LLM behavior using carefully designed malicious prompts and organizes the research into seven key areas.

\textbf{Objective Manipulation:} \hypertarget{ObjectiveManipulation}{} \citet{abdelnabi2023not} demonstrate a prompt injection attack capable of fully compromising LLMs, with practical feasibility showcased on applications like Bing Chat and Github Copilot. \citet{perez2022ignore} introduce the PromptInject framework for goal-hijacking attacks, revealing vulnerability to prompt misalignment and offering insights into inhibiting measures such as stop sequences and post-processing model results.

\textbf{Prompt Leaking:} \citet{liu2023prompt} addresses security vulnerabilities in Large Language Models like GPT-4, focusing on prompt injection attacks. It introduces the HOUYI methodology, a black-box prompt injection attack approach designed for versatility and adaptability across various LLM-integrated services/applications. HOUYI comprises three phases: Context Inference (interaction with the target application to grasp its inherent context and input-output relationships), Payload Generation (devising a prompt generation plan based on the obtained application context and prompt injection guidelines), and Feedback (gauging the effectiveness of the attack by scrutinizing the LLM's responses to the injected prompts, followed by iterative refinement for optimal outcomes), aiming to trick LLMs into interpreting malicious payloads as questions rather than data payloads. Experiments on 36 real-world LLM-integrated services using HOUYI show an 86.1\% success rate in launching attacks, revealing severe ramifications such as unauthorized imitation of services and exploitation of computational power.


\textbf{Malicious Content Generation:} Addressing scalability challenges in malicious prompt generation, \citet{liu2023autodan} present AutoDAN, which is designed to preserve meaningfulness and fluency in prompts. 
They highlight the discovery of prompt injection attacks combined with malicious questions, can lead LLMs to generate harmful or objectionable content by bypassing safety features.  Using a hierarchical genetic algorithm tailored for structured discrete data sets AutoDAN apart from existing methods. The initialization of the population is crucial, and the paper employs handcrafted jailbreak prompts identified by LLM users as prototypes to reduce the search space. Different crossover policies for both sentences and words are introduced to avoid falling into local optima and consistently search for the global optimal solution. Implementation details include a multi-point crossover policy based on a roulette selection strategy and a momentum word scoring scheme to enhance search capability in the fine-grained space. The method achieves lower sentence perplexity, indicating more semantically meaningful and stealthy attacks. 

\textbf{Manipulating Training Data:}\hypertarget{TrainingData}{} \citet{zhao2023prompt} present ProAttack, which boasts near-perfect success rates in evading defenses, highlighting the urgency for better handling of prompt injection attacks with LLMs' growing application.

\textbf{Prompt Injection Attacks and Defenses in LLM-Integrated Applications:} Comprehensive studies such as \cite{liu2023promptattackdefense} emphasize the importance of understanding and mitigating the risks posed by prompt injection attacks. These works highlight sophisticated methodologies like `HouYi' \cite{liu2023promptattackdefense} and underscore the urgent need for more robust security measures.

\textbf{Prompt Manipulation Frameworks:} Recent literature explores various methods for manipulating LLM behavior, as detailed in works like \cite{melamed2023propane,jiang2023prompt}. Propane \cite{melamed2023propane} introduces an automatic prompt optimization framework, while Prompt Packer \cite{jiang2023prompt} introduces Compositional Instruction Attacks, revealing vulnerabilities in LLMs to multifaceted attacks.

\textbf{Benchmarking and Analyzing LLM Prompt Injection Attacks:} \citet{toyer2023tensor} present a dataset of prompt injection attacks and defenses, offering insights into LLM vulnerabilities and paving the way for more resilient systems. This benchmarking and analysis are crucial for understanding the intricacies of prompt injection attacks and developing effective countermeasures.

\subsection{Data Poisoning}

Contemporary NLP systems follow a two-stage process: pretraining and fine-tuning. Pretraining involves learning from a large corpus to understand general linguistic structures, while fine-tuning tailors the model for specific tasks using smaller datasets. Recently, providers like OpenAI have enabled end-users to fine-tune models, enhancing adaptability. This section explores studies on data poisoning techniques and their impact on safety aspects during training, including privacy risks and susceptibility to adversarial attacks.


\textbf{PII extraction}: \citet{chen2023janus} investigate whether fine-tuning large language models (LLMs) on small datasets containing personal identifiable information (PII) can lead to the models disclosing more PII embedded in their original training data. The authors demonstrate a strawman method where an LLM is fine-tuned on a small PII dataset converted to text, which enables the model to then disclose more PII when prompted. To improve on this, they propose Janus methodology which defines a PII recovery task and uses few-shot fine-tuning. Experiments indicate that fine-tuning GPT-3.5 on just 10 PII instances enables it to accurately disclose 650 out of 1000 target PIIs, versus 0 without fine-tuning. The Janus method further improves this divulgence, disclosing 699 target PIIs. Analysis shows larger models and real training data have stronger memorization and PII recovery and fine-tuning is more effective than prompt engineering alone for PII leakage. This indicates that LLMs can shift from non-disclosure to revealing significant amounts of PII with minimal fine-tuning.

\textbf{Bypassing Safety Alignment}: \hypertarget{safety}{} \citet{qi2023fine} investigate safety risks in fine-tuning aligned LLMs, finding that even benign datasets can compromise safety. Backdoor attacks are shown to effectively bypass safety measures, emphasizing the need for improved post-training protections.

\citet{bianchi2023safety} analyze the safety risks of instruction tuning, showing that overly instruction-tuned models can still produce harmful content. They propose a safety tuning dataset to mitigate these risks, balancing safety and model performance.

\citet{zhao2023learning} study how LLMs learn and forget unsafe examples during fine-tuning, proposing a technique called ForgetFilter to filter fine-tuning data and improve safety without sacrificing performance.
\textbf{Backdoor Attacks}:\hypertarget{backdoor}{} \citet{shah2023loft} introduce Local Fine Tuning (LoFT) for discovering adversarial prompts, demonstrating successful attacks on LLMs. \citet{shu2023exploitability} propose Autopoison, an automated data poisoning pipeline, showcasing its effectiveness in altering model behavior without semantic degradation.

\section{Human Interference}

Adversarial attacks range from human-crafted (slow and non-scalable) to automated methods utilizing search, target function optimization, LLMs, and tools.


\subsection{Human Red Teaming}\label{sec:RedTeaming}
Through human-crafted adversarial prompts, individuals employ their creativity and expertise to design attacks carefully. These attacks often involve a deep understanding of the targeted model's vulnerabilities and limitations. 

In a study performed by ~\citet{huang2023catastrophic}, 600 curated harmful prompts were tested over 11 LLMs. By simply varying decoding hyperparameters and sampling methods, they show these curated prompts can easily break LLMs.~\citet{shen2023anything} collect 6,387 malicious prompts and test them over 13 forbidden scenarios from OpenAI's policy. These were collected through various online sources like reddit, discords, datasets and other public places on the web. They found 2 highly effective prompts that have 99\% attack success on GPT-3.5 and GPT-4. 

~\citet{li2023multi} collected personally identifiable information, like emails and phone numbers, to test if they could extract this data from LLMs. They crafted a multi-step jailbreaking role-playing prompting approach that a human attacker can use to break ChatGPT's ethical constraints and extract private data. The website 

The online platform ~\cite{JailBreakChat} is an active website for gathering jailbreaking prompts through crowdsourcing. Another study conducted by ~\citet{liu2023jailbreaking} utilized this website to analyze 78 malicious prompts. These prompts were categorized into three main classes: Pretending, Attention Shifting, and Privilege Escalation, each further divided into subclasses. In total, they created 10 categories encompassing various types of harmful prompts that broke over 10 OpenAI policies. Other sources of curated adversarial prompts have also surfaced over the web ~\cite{SCBSZ23, JailBreakChat}.

Creating interactive systems to facilitate adversarial sample generation is another way to get human expertise into breaking LLMs. ~\citet{wallace2019trick} created an interactive UI for leveraging human creativity and trivia knowledge to generate adversarial examples for Question Answering systems. The authors build an interactive interface that shows question authors model predictions and word importance scores. The authors are trivia enthusiasts who craft tricky questions that fool the model. A similar large scale study ~\cite{schulhoff2023ignore} collect over 600k adversarial prompts from thousands of participants worldwide through an interactive interface. In another work ~\citet{ziegler2022adversarial} leveraged human contractors who manually wrote adversarial text snippets that could fool a injurious/non-injurious text classifier. They built an interface to help contractors rewrite snippets to be adversarial, including highlighting salient tokens and suggesting token replacements. 


~\citet{xu2021bot} introduce Bot-Adversarial Dialogue, a human-and-model-in-the-loop framework for enhancing conversational AI safety. Crowd workers converse with chatbots to elicit unsafe/offensive responses, categorized by severity. A verification task identifies offensive language types, involving humans in both collecting and labeling adversarial examples for safety and offensiveness type.

\subsection{Automated Adversarial Attacks}
Automated adversarial attacks use algorithms to generate and deploy adversarial examples, offering scalability without human expertise.

~\citet{deng2023jailbreaker} propose the ``MASTERKEY framework", which uses time-based characteristics inherent to the generative process to reverse-engineer the defense strategies behind mainstream LLM chatbot services. They automatically generate jailbreak prompts against well-protected LLMs by fine-tuning another LLM with the jailbreak prompts. \citet{zou2023universal} propose a universal automated approach for adversarial attacks on LLMs. It involves generating a suffix to be added to various queries, prompting the LLM to produce inappropriate content. This method merges greedy and gradient-based search techniques to automatically create these adversarial suffixes. The adversarial prompts produced by this method are highly transferable, even to black-box, publicly available, production LLMs.

AutoDAN ~\cite{liu2023autodan}, an automated, interpretable, gradient-based adversarial attack method for LLMs, combines the strengths of manual jailbreak attacks and automatic adversarial attacks. It generates readable prompts that bypass perplexity filters while maintaining high attack success rates. It formulates the attack as an optimization problem and employs a hierarchical genetic algorithm to search for effective prompts in the space initialized by handcrafted prompts. Their method operates at multiple levels - sentence and word - to ensure both diversity and fine-grained optimization.


~\citet{jones2023automatically} present ARCA, a coordinate ascent discrete optimization algorithm efficiently searching for input output text pairs matching a desired behavior in LLMs. It uncovers unexpected behaviors like derogatory completions or language-switching inputs. Several tools to automatically generate adversarial samples for LLMs exist. 
PromptAttack ~\cite{xu2023llm}, a tool for evaluating the adversarial robustness of LLMs, converts adversarial textual attacks into an attack prompt that causes the LLM to output an adversarial sample, essentially fooling itself. The attack prompt consists of the original input, the attack objective, and the attack guidance.

~\citet{casper2023explore} present a red-teaming framework for LLMs, starting with output exploration via clustering, establishing undesired behaviors through classifier training, and using reinforcement learning to train a ``red'' model generating adversarial prompts, focusing on controversial topics. They successfully red-team GPT-2 for toxic text and GPT-3 for false claims, particularly in controversial political contexts, demonstrating more impactful attacks than traditional methods.

\subsection{Mitigation Strategies}
\label{subsec:Mitigation Strategies}
\hypertarget{mitigation-strategies}{}

Mitigation strategies for protecting LLMs can be broadly divided into two categories based on defense deployment strategy.

\subsubsection{External: Input/Output filtering or Guarding}\hypertarget{IO}{}

In guarding-based mitigation for LLMs, external systems play a crucial role by detecting adversarial inputs (input filtering) or anomalous outputs (output filtering), negating the need for model retraining. Popular tools like OpenChatKit\footnote{\url{https://github.com/togethercomputer/OpenChatKit}} and NeMo-Guardrails \citet{rebedea2023nemo} exemplify this approach, and have been adopted by a number of production-LLM systems. Guarding techniques can further be bifurcated into defenses against gradient-based jailbreaks that employ adversarial suffixes to augment prompts, and manual jailbreaks aiming to misalign the model's responses.

\textbf{Defense against gradient-based jailbreaks:} The current state-of-the-art literature in the area of mitigating gradient-based adversarial attacks on LLMs can be broadly categorized into two main strategies: one focusing on detecting malicious prompts based on characteristic features of the input (e.g., high perplexity, character-level perturbations) and the other utilizing classifier-based approaches where models, such as DistilBERT\cite{sanh2019distilbert}, are employed to distinguish between adversarial and non-adversarial prompts.

In the former category, \citet{jain2023baseline} discuss baseline defenses like input filtering, which, despite their effectiveness, may inadvertently alter the intended output through techniques such as paraphrasing and retokenization, or flag legitimate queries due to perplexity-based filtering. Similarly, \citet{robey2023smoothllm} introduce SmoothLLM, which leverages the vulnerability of adversarial attacks to character-level perturbations, adopting a scatter-gather approach for prompt processing. This method aims to nullify adversarial content by averaging out the final response based on the aggregated responses produced by the model for the perturbed input prompts. Similarly, \citet{hu2023tokenlevel} propose token-level adversarial prompt detection, capitalizing on the high perplexity characteristic of adversarial prompts to identify and classify adversarial tokens within a prompt, leveraging the relationship between neighbouring tokens. As with other perplexity-based techniques, this might not be feasible for black-box LLMs where perplexity calculation cannot be done directly.

On the classifier-based side, \citet{kim2023robust} propose the Adversarial Prompt Shield (APS), a DistilBERT\cite{sanh2019distilbert}-based model designed for prompt classification into safe or unsafe categories. This approach is complemented by a method for generating training data that simulates adversarial attacks by adding synthetic noise to legitimate conversations. However, the necessity for frequent retraining to stay abreast of new attack vectors and reduce false positives presents a challenge to this approach.

The characteristic feature-based methods provide a more direct approach to detecting adversarial content, potentially allowing for real-time mitigation without the need for extensive retraining. Conversely, classifier-based approaches, while requiring more maintenance, offer a more nuanced understanding of the intricacies of adversarial and non-adversarial prompts, potentially leading to more accurate and robust defenses against a wider range of attacks.

\textbf{Defense against manual jailbreaks:} \citet{inan2023llama} introduce Llama Guard, a safeguard model leveraging Llama2-7b \citet{touvron2023llama} for input-output protection in LLMs. It employs taxonomy-based task classification for customizing responses through few-shot prompting or fine-tuning. \citet{rebedea2023nemo} present NeMo-Guardrails, an open-source framework enhancing LLM conversational systems with programmable guardrails. It uses a proxy layer with Colang-defined rules to manage user interactions, though its reliance on chain-of-thought (CoT) prompting may limit scalability. \citet{helbling2023llm} propose a similar approach and suggest an output filtering method involving a secondary LLM to assess the malicious nature of responses, facing challenges in language compatibility and operational costs.

\citet{glukhov2023llm} argue that semantic censorship in LLMs is inherently undecidable, given their ability to follow instructions and generate outputs through arbitrary rule-based encodings. They propose viewing LLM censorship as a security issue, necessitating specific countermeasures rather than treating it solely as a machine learning challenge.

\subsubsection{Internal: Model training/fine-tuning} \hypertarget{MTFT}{}

The state of the art methods in this differ primarily in the stage at which the model is trained for providing safe outputs, as well as the source of the data used for providing the safe output. In this section, we highlight the current trends.

\textbf{Supervised Safety fine-tuning:} \citet{touvron2023llama}, collect adversarial prompts along with their safe demonstrations and then use these samples as a part of the general supervised fine tuning pipeline. While the examples in this case are curated manually, automated collection techniques and red-teaming are an effective methods to discover harmful prompts. A detailed discussion of red-teaming and collecting data both manually and automatically is discussed in \cref{sec:RedTeaming}.

\textbf{Safety-tuning as a part of the RLHF pipeline:} RLHF has been shown to make models more robust to jailbreak attempts \citet{bai2022training}. \citet{touvron2023llama} train a safety reward model based on manually collected adversarial prompts and responses from multiple models where the response that is deemed the safest is selected, this reward model is then used as a part of RLHF pipeline in order to safety-tune the model. 

\textbf{Safety Context Distillation}: In using Context Distillation \citet{askell2021general} for model safety, \citet{touvron2023llama} prepend the prompt with a persona of a safe model such as “You are a responsible and safe assistant,” and then while fine-tuning, they remove this prepended prompt, distilling this safe context into the model, enhancing its proclivity to deny any requests that create a problematic response. 
\section{Challenges and Future Research}
\label{sec:future-research}
\hypertarget{future-research}{}

Here, we discuss a few potential directions that are promising for future research on defending the attacks on LLMs, enhancing their robustness, and gaining trust from the end-users.


\subsection{Real-time Monitoring Systems}
The growing use of Large Language Models (LLMs) in diverse fields brings various applications, but it requires robust monitoring to detect anomalies effectively. Current evaluation mechanisms are inadequate, leaving LLMs vulnerable to threats like data exposure, misinformation, illegal content, and aiding criminal activities. Understanding and countering these attacks are challenging due to adversaries' ability to manipulate LLMs with deceptive prompts. Therefore, it is imperative to not only introduce LLM safeguard systems but also to fortify them with advanced detection capabilities. Future research can focus on building such systems, equipped to scrutinize outputs comprehensively, identifying and flagging any undesirable content swiftly and accurately. Additionally, efforts should be directed towards ensuring the resilience and adaptability of these guard mechanisms, making them resistant to potential evasion tactics employed by adversaries. 

\subsection{Multimodal Approach}

The integration of multimodal capabilities presents both exciting opportunities and formidable challenges for ensuring the safety and reliability of LLMs. Future research should prioritize developing techniques to mitigate these challenges, such as improving input sanitization and validation, and creating custom defense prompts to prevent jailbreaking attempts. These efforts are crucial for strengthening the security and resilience of LLMs amidst evolving threats in multimodal environments.

\subsection{Benchmark}
It becomes apparent that safeguarding LLMs alone falls short of addressing the broader concerns. Hence, the pertinent inquiry emerges: How can we reliably determine the comparative efficacy of attacks `A' versus `B' on LLMs through quantifiable and rational observations? The establishment of a standardized benchmark for evaluating attacks on LLMs becomes important, ensuring ethical reliability and factual performance. While considerable research has been devoted to benchmarking ~\cite{jin2024attackeval}, the existing frameworks often prove insufficient for practical deployment in real-world scenarios. Consequently, the development of a scalable, near-real-time evaluation infrastructure emerges as a crucial requirement for both LLMs and their enterprise applications.

\subsection{Explainable LLMs}

Explainability of LLMs is pivotal, not just for enhancing the transparency and trustworthiness of these models but also for identifying and mitigating vulnerabilities to linguistic attacks. Future research in explainable LLMs must pivot towards developing and refining methods that illuminate the complex decision-making processes inherent within these models. This entails a focused investigation into explainability techniques that unravel the intricacies of attention mechanisms, delineates the significance of features contributing to the models' outputs, and trace the reasoning pathways that underpin their decisions. Such efforts are critical for enabling a deeper understanding and interpretation of LLM outputs by a broad spectrum of stakeholders, from developers to end-users. There are existing work ~\cite{chefer2021transformer, voita2019analyzing, dosovitskiy2020image} that try to explain transformer architecture outputs but because of the black box nature of neural networks, they fall short to give reliable explanations and leave room for further fundamental developments. Moreover, the endeavor to make LLMs explainable presents multifaceted challenges, including the technical difficulty of dissecting often opaque neural network architectures, the need for methodologies that can reliably attribute decision-making in a manner that is both accurate and accessible to non-experts, and the ethical implications of creating transparent systems that respect user privacy and data security. Addressing these challenges requires a multidisciplinary approach that bridges computational techniques with principles of ethical AI, aiming to foster models that are not only robust and efficient but also intrinsically interpretable and aligned with societal values. This push towards explainable LLMs is not just a technical necessity but a foundational step towards ensuring that AI technologies remain accountable, understandable, and beneficial across diverse applications.

\section{Conclusion}
\label{sec:conclusion}

This paper provides a comprehensive overview of attacks targeting LLMs. We start by categorizing the LLM attacks literature into a novel taxonomy to provide a better structure and aid for future research. Through the examination of these attack vectors, it is evident that LLMs are vulnerable to a diverse range of threats, posing significant challenges to their security and reliability in real-world applications. Furthermore, this paper has highlighted the importance of implementing effective mitigation strategies to defend against LLM attacks. These strategies encompass a variety of approaches, including data filtering, guardrails, robust training techniques, adversarial training and safety context distillation. To summarize, although LLMs present significant opportunities for enhancing natural language processing capabilities, their vulnerability to adversarial exploitation highlights the critical need to address security issues. Through ongoing exploration and advancement in detecting attacks, implementing mitigative measures, and enhancing model resilience, we can aim to fully leverage the advantages of LLM technology while fortifying defenses against potential risks.



\section{Limitations}

This study, while comprehensive in its examination of attacks on Large Language Models (LLMs) and mitigation strategies, is subject to several limitations:

\textbf{Scope and Coverage:} Despite our efforts to conduct a thorough survey, the fast-paced advancements in LLM technologies and attack methodologies mean that some emerging threats might not be covered. The landscape of cybersecurity threats evolves rapidly, and new vulnerabilities could emerge following this publication.
    
\textbf{Generalizability of Mitigation Strategies:} The effectiveness of the mitigation strategies discussed may vary across different models, contexts, and against specific attacks. While we aimed for broad applicability in our recommendations, the specificity of certain defenses to particular models or scenarios limits their universal applicability.
    
\textbf{Ethical and Societal Implications:} Our focus was primarily on the technical aspects of LLM security, which led to a less comprehensive exploration of the broader ethical and societal implications of both the attacks and the countermeasures. The dual-use nature of many AI technologies, including those discussed, necessitates careful consideration of ethical implications beyond the scope of this paper.
    
\textbf{Dynamic Nature of Threats:} The adversarial landscape is characterized by an ongoing race, with attackers continually evolving their strategies in response to new defenses. This paper captures a snapshot of the current state, but continuous research and vigilance are required to address the adaptive nature of threats.

\textbf{Scalability and Practicality of Defenses:} Implementing robust defense mechanisms in practical settings poses challenges, including computational overhead, scalability issues, and the need for ongoing updates. Balancing security with usability remains a critical, yet underexplored, area.

In summary, while this work provides significant insights into LLM security, it highlights the importance of continued research, interdisciplinary collaboration, and an agile response to the complex and evolving landscape of AI security.

\bibliography{anthology,custom}
\bibliographystyle{acl_natbib}

\newpage
\onecolumn
\appendix

\section{Appendix}
\label{sec:appendix}

\begin{table}[htbp]
\centering
\scalebox{0.68}{
\begin{tabular}{lccccccl}
\toprule
\multicolumn{1}{c}{\multirow{3}{*}{\textbf{Paper Name}}} & \multicolumn{1}{c}{\multirow{3}{*}{\textbf{Category}}} & \multicolumn{3}{c}{{\textbf{Threat Model}}} & \multicolumn{2}{c}{\textbf{Attack Strategy}} & \multicolumn{1}{c}{\multirow{3}{*}{\textbf{\makecell{Evaluated\\LLM Models}}}} \\
\cmidrule{3-5} \cmidrule{6-7}
 &  & \multirow{2}{*}{\textbf{\makecell{White\\Box}}} & \multirow{2}{*}{\textbf{\makecell{Gray\\Box}}} & \multirow{2}{*}{\textbf{\makecell{Black\\Box}}} & \multirow{2}{*}{\textbf{\makecell{Prompt or\\Response}}} & \multirow{2}{*}{\textbf{\makecell{Model\\Based}}} &  \\
 &  &  &  &  &  &  &  \\
\midrule
\cite{chao2023jailbreaking} & Jailbreak & & & \cmark & \cmark & & GPT-3.5/4, Vicuna, and PaLM-2 \\
\cite{wei2023jailbroken} & Jailbreak & & & \cmark & \cmark & & GPT-4, GPT-3.5 Turbo, Claude v1.3 \\
\cite{li2023deepinception} & Jailbreak & & & \cmark & \cmark & \cmark & Falcon, Vicuna, Llama-2, GPT-3.5, GPT-4, GPT-4V \\
\cite{ding2023wolf} & Jailbreak & & & \cmark & \cmark & \cmark & ChatGPT, GPT-4 \\
\cite{qi2023visual} & Jailbreak & \cmark & & & \cmark & \cmark & MiniGPT-4, BLIP-2, GPT-4 \\
\cite{yong2023low} & Jailbreak & & & \cmark & \cmark & & GPT-4 \\
\cite{zou2023universal} & Jailbreak & \cmark & & \cmark & \cmark & \cmark & Vicuna, ChatGPT, Claude, Llama-2, Pythia, Falcon \\
\cite{mehrotra2023tree} & Jailbreak & & & \cmark & \cmark & \cmark & GPT-4, GPT-4 Turbo \\
\hline
\cite{abdelnabi2023not} & Prompt Injection & & \cmark & \cmark & \cmark & & GPT-3.5, GPT-4 \\
\cite{perez2022ignore} & Prompt Injection & & & \cmark & \cmark & & GPT-3.5 \\
\cite{zhao2023prompt} & Prompt Injection & \cmark & & \cmark &  & \cmark & GPT-NEO \\
\cite{liu2023promptattackdefense} & Prompt Injection & & \cmark & \cmark & \cmark & \cmark & GPT-3.5 \\
\cite{toyer2023tensor} & Prompt Injection & & \cmark & & \cmark &  & Llama-2 (7B, 13B, 70B), CodeLaMMA-34B \\
\cite{melamed2023propane} & Prompt Injection & & & \cmark &  &  & GPT Model suits: Pythia \\
\cite{jiang2023prompt} & Prompt Injection & & & \cmark & \cmark &  & GPT-4, GPT-3.5, and ChatGLM2-6B \\
\hline
\cite{chen2023janus} & Data Poisoning & &  & \cmark & \cmark & & GPT-3.5 \\
\cite{lukas2023analyzing} & Data Poisoning & &\cmark & \cmark & \cmark & & GPT-3.5 \\
\cite{qi2023fine} & Data Poisoning & \cmark & & \cmark &  & \cmark & GPT-3.5 Turbo, Llama-2 \\
\cite{zhan2023removing} & Data Poisoning & & \cmark &\cmark  & \cmark &  & GPT-4 \\
\cite{bianchi2023safety} & Data Poisoning & & \cmark & & \cmark &  & LLaMA, Falcon \\
\cite{zhao2023learning} & Data Poisoning & & & \cmark &  &  & LLaMA 7B \\
\cite{shah2023loft} & Data Poisoning & & & \cmark & \cmark &  & ChatGPT, GPT-4, and Claude \\
\cite{shu2023exploitability} & Data Poisoning & & & \cmark &\cmark  &  & OPT (350M, 1.3B, 6.7B) \\

\bottomrule
\end{tabular}
}
\caption{A comprehensive summary detailing attacks targeting LLMs is provided, categorized into three primary categories: Jailbreak, Prompt Injection, and Data Poisoning. We outline the threat model, attack strategy, and the list of evaluated LLM models for each of the papers listed.}
\label{tab:attacks_summary}
\end{table}

\end{document}